\documentclass[11pt]{cernrep}
\usepackage{graphicx}
\usepackage{epsfig}

\begin{document}


\newcommand{\invfb}{\ensuremath{\mathrm{fb^{-1}}}}
\newcommand{\invpb}{\ensuremath{\mathrm{pb^{-1}}}}
\newcommand{\invnb}{\ensuremath{\mathrm{nb^{-1}}}}
\newcommand{\rhoprim}{\mbox{$\rho^\prime$}}
\newcommand{\tprim}{\mbox{$t^\prime$}}
\newcommand{\rhop}{\mbox{$\rho^\prime$}}
\newcommand{\zet}{\mbox{$\zeta$}}
\newcommand{\rfivecomb}{\mbox{$r^5_{00} + 2 r^5_{11}$}}
\newcommand{\ronecomb}{\mbox{$r^1_{00} + 2 r^1_{11}$}}
\newcommand{\gstarVM} {\mbox{$\gamma^* \ \!p \rightarrow V\ \!Y$}}
\newcommand{\gsrel} {\mbox{$\gamma^* \ \!p \rightarrow \rho \ \!p$}}
\newcommand{\gsrpd} {\mbox{$\gamma^* \ \!p \rightarrow \rho \ \!Y$}}
\newcommand{\gspel} {\mbox{$\gamma^* \ \!p \rightarrow \phi \ \!p$}}
\newcommand{\gsppd} {\mbox{$\gamma^* \ \!p \rightarrow \phi \ \!Y$}}
\newcommand{\gstarp} {\mbox{$\gamma^*\ \!p$}}
\newcommand{\mv} {\mbox{$M_V$}}
\newcommand{\mvsq} {\mbox{$M_V^2$}}
\newcommand{\msq} {\mbox{$M_V^2$}}
\newcommand{\qsqplmsq} {\mbox{($Q^2 \!+ \!M_V^2$})}
\newcommand{\qqsqplmsq} {\mbox{$Q^2 \!+ \!M_V^2}$}
\newcommand{\alprim}{\mbox{$\alpha^\prime$}}
\newcommand{\alphaz}{\mbox{$\alpha(0)$}}
\newcommand{\alpomz}{\mbox{$\alpha_{\PO}(0)$}}
\newcommand{\hence}{\mbox{$=>$}}
\newcommand{\vm}{\mbox{$V\!M$}}
\newcommand{\sur}{\mbox{\ \! / \ \!}}
\newcommand{\tzz} {\mbox{$T_{00}$}}
\newcommand{\tuu} {\mbox{$T_{11}$}}
\newcommand{\tzu} {\mbox{$T_{01}$}}
\newcommand{\tuz} {\mbox{$T_{10}$}}
\newcommand{\tmuu} {\mbox{$T_{-11}$}}
\newcommand{\ralpha} {\mbox{$\tuu \sur \tzz$}}
\newcommand{\rbeta} {\mbox{$\tzu \sur \tzz$}}
\newcommand{\rdelta} {\mbox{$\tuz \sur \tzz$}}
\newcommand{\reta} {\mbox{$\tmuu \sur \tzz$}}
\newcommand{\averm} {\mbox{$\av {M}$}}
\newcommand{\rapproch} {\mbox{$R_{SCHC+T_{01}}$}}
\newcommand{\chisq} {\mbox{$\chi^2 / {\rm d.o.f.}$}}


%
%
\newcommand{\s}{\mbox{$s$}}
\newcommand{\ttra}{\mbox{$t$}}
\newcommand{\modt}{\mbox{$|t|$}}
\newcommand{\eminpz}{\mbox{$E-p_z$}}
\newcommand{\eminpzs}{\mbox{$\Sigma(E-p_z)$}}
\newcommand{\rap}{\ensuremath{\eta^*} }
\newcommand{\W}{\mbox{$W$}}
\newcommand{\w}{\mbox{$W$}}
\newcommand{\Q}{\mbox{$Q$}}
\newcommand{\q}{\mbox{$Q$}}
\newcommand{\xB}{\mbox{$x$}}  
\newcommand{\xF}{\mbox{$x_F$}}  
\newcommand{\xg}{\mbox{$x_g$}}  
\newcommand{\xbj}{x}
\newcommand{\xpom}{x_{\PO}}
\newcommand{\y}{\mbox{$y~$}}
\newcommand{\Qsq}{\mbox{$Q^2$}}
\newcommand{\qsq}{\mbox{$Q^2$}}
\newcommand{\kjet}{\mbox{$k_{T\rm{jet}}$}}
\newcommand{\xjet}{\mbox{$x_{\rm{jet}}$}}
\newcommand{\Ejet}{\mbox{$E_{\rm{jet}}$}}
\newcommand{\thjet}{\mbox{$\theta_{\rm{jet}}$}}
\newcommand{\pjet}{\mbox{$p_{T\rm{jet}}$}}
\newcommand{\et}{\mbox{$E_T~$}}
\newcommand{\kt}{\mbox{$k_T~$}}
\newcommand{\ptrans}{\mbox{$p_T~$}}
\newcommand{\pth}{\mbox{$p_T^h~$}}
\newcommand{\pte}{\mbox{$p_T^e~$}}
\newcommand{\ptsq}{\mbox{$p_T^{\star 2}~$}}
\newcommand{\as}{\mbox{$\alpha_s~$}}
\newcommand{\ycut}{\mbox{$y_{\rm cut}~$}}
\newcommand{\gx}{\mbox{$g(x_g,Q^2)$~}}
\newcommand{\xpart}{\mbox{$x_{\rm part~}$}}
\newcommand{\mrsdm}{\mbox{${\rm MRSD}^-~$}}
\newcommand{\mrsdmp}{\mbox{${\rm MRSD}^{-'}~$}}
\newcommand{\mrsdn}{\mbox{${\rm MRSD}^0~$}}
\newcommand{\lambdams}{\mbox{$\Lambda_{\rm \bar{MS}}~$}}
%
%
\newcommand{\gp}{\ensuremath{\gamma}p }
\newcommand{\gammasp}{\ensuremath{\gamma}*p }
\newcommand{\gammap}{\ensuremath{\gamma}p }
\newcommand{\dsiget}{\ensuremath{{\rm d}\sigma_{ep}/{\rm d}E_t^*} }
\newcommand{\dsigrap}{\ensuremath{{\rm d}\sigma_{ep}/{\rm d}\eta^*} }
\newcommand{\epem}{\mbox{$e^+e^-$}}
\newcommand{\ep}{\mbox{$ep~$}}
\newcommand{\epl}{\mbox{$e^{+}$}}
\newcommand{\emi}{\mbox{$e^{-}$}}
\newcommand{\epm}{\mbox{$e^{\pm}$}}
\newcommand{\se}{section efficace}
\newcommand{\ses}{sections efficaces}
%
%
\newcommand{\phib}{\mbox{$\varphi$}}
\newcommand{\rh}{\mbox{$\rho$}}
\newcommand{\rhz}{\mbox{$\rh^0$}}
\newcommand{\ph}{\mbox{$\phi$}}
\newcommand{\om}{\mbox{$\omega$}}
\newcommand{\jpsi}{\mbox{$J/\psi$}}
\newcommand{\pipi}{\mbox{$\pi^+\pi^-$}}
\newcommand{\pip}{\mbox{$\pi^+$}}
\newcommand{\pim}{\mbox{$\pi^-$}}
\newcommand{\kk}{\mbox{K^+K^-$}}
\newcommand{\bsl}{\mbox{$b$}}
\newcommand{\alp}{\mbox{$\alpha^\prime$}}
\newcommand{\alpom}{\mbox{$\alpha_{\PO}$}}
\newcommand{\alpomp}{\mbox{$\alpha_{\PO}^\prime$}}
\newcommand{\rzzzz}{\mbox{$r_{00}^{04}$}}
\newcommand{\rzqzz}{\mbox{$r_{00}^{04}$}}
\newcommand{\rzquz}{\mbox{$r_{10}^{04}$}}
\newcommand{\rzqumu}{\mbox{$r_{1-1}^{04}$}}
\newcommand{\ruuu}{\mbox{$r_{11}^{1}$}}
\newcommand{\ruzz}{\mbox{$r_{00}^{1}$}}
\newcommand{\ruuz}{\mbox{$r_{10}^{1}$}}
\newcommand{\ruumu}{\mbox{$r_{1-1}^{1}$}}
\newcommand{\rduz}{\mbox{$r_{10}^{2}$}}
\newcommand{\rdumu}{\mbox{$r_{1-1}^{2}$}}
\newcommand{\rcuu}{\mbox{$r_{11}^{5}$}}
\newcommand{\rczz}{\mbox{$r_{00}^{5}$}}
\newcommand{\rcuz}{\mbox{$r_{10}^{5}$}}
\newcommand{\rcumu}{\mbox{$r_{1-1}^{5}$}}
\newcommand{\rsuz}{\mbox{$r_{10}^{6}$}}
\newcommand{\rsumu}{\mbox{$r_{1-1}^{6}$}}
\newcommand{\rzqik}{\mbox{$r_{ik}^{04}$}}
\newcommand{\rhzik}{\mbox{$\rh_{ik}^{0}$}}
\newcommand{\rhqik}{\mbox{$\rh_{ik}^{4}$}}
\newcommand{\rhaik}{\mbox{$\rh_{ik}^{\alpha}$}}
\newcommand{\rhzzz}{\mbox{$\rh_{00}^{0}$}}
\newcommand{\rhqzz}{\mbox{$\rh_{00}^{4}$}}
\newcommand{\raik}{\mbox{$r_{ik}^{\alpha}$}}
\newcommand{\razz}{\mbox{$r_{00}^{\alpha}$}}
\newcommand{\rauz}{\mbox{$r_{10}^{\alpha}$}}
\newcommand{\raumu}{\mbox{$r_{1-1}^{\alpha}$}}

\newcommand{\R}{\mbox{$R$}}
\newcommand{\rzero}{\mbox{$r_{00}^{04}$}}
\newcommand{\rone}{\mbox{$r_{1-1}^{1}$}}
\newcommand{\costh}{\mbox{$\cos\theta$}}
\newcommand{\cosp}{\mbox{$\cos\psi$}}
\newcommand{\costop}{\mbox{$\cos(2\psi)$}}
\newcommand{\cosd}{\mbox{$\cos\delta$}}
\newcommand{\cossqp}{\mbox{$\cos^2\psi$}}
\newcommand{\cossqt}{\mbox{$\cos^2\theta^*$}}
\newcommand{\sint}{\mbox{$\sin\theta^*$}}
\newcommand{\sintot}{\mbox{$\sin(2\theta^*)$}}
\newcommand{\sinsqt}{\mbox{$\sin^2\theta^*$}}
\newcommand{\costhst}{\mbox{$\cos\theta^*$}}
\newcommand{\vep}{\mbox{$V p$}}
\newcommand{\mpipi}{\mbox{$m_{\pi^+\pi^-}$}}
\newcommand{\mkk}{\mbox{$m_{KK}$}}
\newcommand{\mkaka}{\mbox{$m_{K^+K^-}$}}
\newcommand{\mpp}{\mbox{$m_{\pi\pi}$}}       
\newcommand{\mppsq}{\mbox{$m_{\pi\pi}^2$}}   
\newcommand{\mpi}{\mbox{$m_{\pi}$}}          
\newcommand{\mrho}{\mbox{$m_{\rho}$}}        
\newcommand{\mrhosq}{\mbox{$m_{\rho}^2$}}    
\newcommand{\Gmpp}{\mbox{$\Gamma (\mpp)$}}   
\newcommand{\Gmppsq}{\mbox{$\Gamma^2(\mpp)$}}
\newcommand{\Grho}{\mbox{$\Gamma_{\rho}$}}   
\newcommand{\grho}{\mbox{$\Gamma_{\rho}$}}   
\newcommand{\Grhosq}{\mbox{$\Gamma_{\rho}^2$}}   
%
%
\newcommand{\cm}{\mbox{\rm cm}}
\newcommand{\GeV}{\mbox{\rm GeV}}
\newcommand{\gev}{\mbox{\rm GeV}}
\newcommand{\GeVx}{\rm GeV}
\newcommand{\gevx}{\rm GeV}
\newcommand{\GeVc}{\rm GeV/c}
\newcommand{\gevc}{\rm GeV/c}
\newcommand{\MeVc}{\rm MeV/c}
\newcommand{\mevc}{\rm MeV/c}
\newcommand{\MeV}{\mbox{\rm MeV}}
\newcommand{\mev}{\mbox{\rm MeV}}
\newcommand{\MeVx}{\mbox{\rm MeV}}
\newcommand{\mevx}{\mbox{\rm MeV}}
\newcommand{\GeVsq}{\mbox{${\rm GeV}^2$}}
\newcommand{\gevsq}{\mbox{${\rm GeV}^2$}}
\newcommand{\gevsqc}{\mbox{${\rm GeV^2/c^4}$}}
\newcommand{\gevcsq}{\mbox{${\rm GeV/c^2}$}}
\newcommand{\mevcsq}{\mbox{${\rm MeV/c^2}$}}
\newcommand{\GeVsqm}{\mbox{${\rm GeV}^{-2}$}}
\newcommand{\gevsqm}{\mbox{${\rm GeV}^{-2}$}}
\newcommand{\nb}{\mbox{${\rm nb}$}}
\newcommand{\nbinv}{\mbox{${\rm nb^{-1}}$}}
\newcommand{\pbinv}{\mbox{${\rm pb^{-1}}$}}
\newcommand{\mm}{\mbox{$\cdot 10^{-2}$}}
\newcommand{\mmm}{\mbox{$\cdot 10^{-3}$}}
\newcommand{\mmmm}{\mbox{$\cdot 10^{-4}$}}
\newcommand{\degr}{\mbox{$^{\circ}$}}
%
%
\newcommand{\F}{$ F_{2}(x,Q^2)\,$}  
\newcommand{\Fc}{$ F_{2}\,$}    
\newcommand{\XP}{x_{{I\!\!P}/{p}}}       
\newcommand{\TOSS}{x_{{i}/{\PO}}}        
\newcommand{\un}[1]{\mbox{\rm #1}} 
\newcommand{\LO}{Leading Order}
\newcommand{\NLO}{Next to Leading Order}
\newcommand{\ft}{$ F_{2}\,$}
%
%
\newcommand{\mc}{\multicolumn}
\newcommand{\bce}{\begin{center}}
\newcommand{\ece}{\end{center}}
\newcommand{\beq}{\begin{equation}}
\newcommand{\eeq}{\end{equation}}
\newcommand{\bea}{\begin{eqnarray}}
\newcommand{\eea}{\end{eqnarray}}
%
%
\def\lsim{\mathrel{\rlap{\lower4pt\hbox{\hskip1pt$\sim$}}
    \raise1pt\hbox{$<$}}}         
\def\gsim{\mathrel{\rlap{\lower4pt\hbox{\hskip1pt$\sim$}}
    \raise1pt\hbox{$>$}}}         
%
%
\newcommand{\pom}{{I\!\!P}}
\newcommand{\PO}{I\!\!P}
\newcommand{\slowpi}{\pi_{\mathit{slow}}}
\newcommand{\fiidiii}{F_2^{D(3)}}
\newcommand{\fiidiiiarg}{\fiidiii\,(\beta,\,Q^2,\,x)}
\newcommand{\n}{1.19\pm 0.06 (stat.) \pm0.07 (syst.)}
\newcommand{\nz}{1.30\pm 0.08 (stat.)^{+0.08}_{-0.14} (syst.)}
\newcommand{\fiidiiiful}{F_2^{D(4)}\,(\beta,\,Q^2,\,x,\,t)}
\newcommand{\fiipom}{\tilde F_2^D}
\newcommand{\ALPHA}{1.10\pm0.03 (stat.) \pm0.04 (syst.)}
\newcommand{\ALPHAZ}{1.15\pm0.04 (stat.)^{+0.04}_{-0.07} (syst.)}
\newcommand{\fiipomarg}{\fiipom\,(\beta,\,Q^2)}
\newcommand{\pomflux}{f_{\pom / p}}
\newcommand{\nxpom}{1.19\pm 0.06 (stat.) \pm0.07 (syst.)}
\newcommand {\gapprox}
   {\raisebox{-0.7ex}{$\stackrel {\textstyle>}{\sim}$}}
\newcommand {\lapprox}
   {\raisebox{-0.7ex}{$\stackrel {\textstyle<}{\sim}$}}
\newcommand{\pomfluxarg}{f_{\pom / p}\,(x_\pom)}
\newcommand{\dsf}{\mbox{$F_2^{D(3)}$}}
\newcommand{\dsfva}{\mbox{$F_2^{D(3)}(\beta,Q^2,x_{I\!\!P})$}}
\newcommand{\dsfvb}{\mbox{$F_2^{D(3)}(\beta,Q^2,x)$}}
\newcommand{\dsfpom}{$F_2^{I\!\!P}$}
\newcommand{\gap}{\stackrel{>}{\sim}}
\newcommand{\lap}{\stackrel{<}{\sim}}
\newcommand{\fem}{$F_2^{em}$}
\newcommand{\tsnmp}{$\tilde{\sigma}_{NC}(e^{\mp})$}
\newcommand{\tsnm}{$\tilde{\sigma}_{NC}(e^-)$}
\newcommand{\tsnp}{$\tilde{\sigma}_{NC}(e^+)$}
\newcommand{\st}{$\star$}
\newcommand{\sst}{$\star \star$}
\newcommand{\ssst}{$\star \star \star$}
\newcommand{\sssst}{$\star \star \star \star$}
\newcommand{\tw}{\theta_W}
\newcommand{\sw}{\sin{\theta_W}}
\newcommand{\cw}{\cos{\theta_W}}
\newcommand{\sww}{\sin^2{\theta_W}}
\newcommand{\cww}{\cos^2{\theta_W}}
\newcommand{\trm}{m_{\perp}}
\newcommand{\trp}{p_{\perp}}
\newcommand{\trmm}{m_{\perp}^2}
\newcommand{\trpp}{p_{\perp}^2}
\newcommand{\ev}{\'ev\'enement}
\newcommand{\evs}{\'ev\'enements}
\newcommand{\mdv}{mod\`ele \`a dominance m\'esovectorielle}
\newcommand{\mdmv}{mod\`ele \`a dominance m\'esovectorielle}
\newcommand{\mdm}{mod\`ele \`a dominance m\'esovectorielle}
\newcommand{\idiff}{interaction diffractive}
\newcommand{\idiffs}{interactions diffractives}
\newcommand{\pdmv}{production diffractive de m\'esons vecteurs}
\newcommand{\pdmr}{production diffractive de m\'esons \rh}
\newcommand{\pdmp}{production diffractive de m\'esons \ph}
\newcommand{\pdmo}{production diffractive de m\'esons \om}
\newcommand{\pdm}{production diffractive de m\'esons}
\newcommand{\pdiff}{production diffractive}
\newcommand{\diff}{diffractive}
\newcommand{\produ}{production}
\newcommand{\mvs}{m\'esons vecteurs}
\newcommand{\me}{m\'eson}
\newcommand{\mr}{m\'eson \rh}
\newcommand{\mph}{m\'eson \ph}
\newcommand{\mo}{m\'eson \om}
\newcommand{\mrs}{m\'esons \rh}
\newcommand{\mps}{m\'esons \ph}
\newcommand{\mos}{m\'esons \om}
\newcommand{\photo}{photoproduction}
\newcommand{\agq}{\`a grand \qsq}
\newcommand{\agqsq}{\`a grand \qsq}
\newcommand{\apq}{\`a petit \qsq}
\newcommand{\apqsq}{\`a petit \qsq}
\newcommand{\de}{d\'etecteur}
%
%
\newcommand{\sqrts}{$\sqrt{s}$}
\newcommand{\Oa}{$O(\alpha_s)$}
\newcommand{\Oaa}{$O(\alpha_s^2)$}
\newcommand{\PT}{p_{\perp}}
\newcommand{\sh}{\hat{s}}
\newcommand{\uh}{\hat{u}}
\newcommand{\ttbs}{\char'134}
\newcommand{\xpomlo}{3\times10^{-4}}
\newcommand{\xpomup}{0.05}
\newcommand{\llq}{$\alpha_s \ln{(\qsq / \Lambda_{QCD}^2)}$}
\newcommand{\llqx}{$\alpha_s \ln{(\qsq / \Lambda_{QCD}^2)} \ln{(1/x)}$}
\newcommand{\llx}{$\alpha_s \ln{(1/x)}$}
%
%
\newcommand{\Brodsky}{Brodsky {\it et al.}}
\newcommand{\FKS}{Frankfurt, Koepf and Strikman}
\newcommand{\Kop}{Kopeliovich {\it et al.}}
\newcommand{\Ginzburg}{Ginzburg {\it et al.}}
\newcommand{\Ryskin}{\mbox{Ryskin}}
\newcommand{\Kaidalov}{Kaidalov {\it et al.}}
%
%
\def\ar#1#2#3   {{\em Ann. Rev. Nucl. Part. Sci.} {\bf#1} (#2) #3}
\def\epj#1#2#3  {{\em Eur. Phys. J.} {\bf#1} (#2) #3}
\def\err#1#2#3  {{\it Erratum} {\bf#1} (#2) #3}
\def\ib#1#2#3   {{\it ibid.} {\bf#1} (#2) #3}
\def\ijmp#1#2#3 {{\em Int. J. Mod. Phys.} {\bf#1} (#2) #3}
\def\jetp#1#2#3 {{\em JETP Lett.} {\bf#1} (#2) #3}
\def\mpl#1#2#3  {{\em Mod. Phys. Lett.} {\bf#1} (#2) #3}
\def\nim#1#2#3  {{\em Nucl. Instr. Meth.} {\bf#1} (#2) #3}
\def\nc#1#2#3   {{\em Nuovo Cim.} {\bf#1} (#2) #3}
\def\np#1#2#3   {{\em Nucl. Phys.} {\bf#1} (#2) #3}
\def\pl#1#2#3   {{\em Phys. Lett.} {\bf#1} (#2) #3}
\def\prep#1#2#3 {{\em Phys. Rep.} {\bf#1} (#2) #3}
\def\prev#1#2#3 {{\em Phys. Rev.} {\bf#1} (#2) #3}
\def\prl#1#2#3  {{\em Phys. Rev. Lett.} {\bf#1} (#2) #3}
\def\ptp#1#2#3  {{\em Prog. Th. Phys.} {\bf#1} (#2) #3}
\def\rmp#1#2#3  {{\em Rev. Mod. Phys.} {\bf#1} (#2) #3}
\def\rpp#1#2#3  {{\em Rep. Prog. Phys.} {\bf#1} (#2) #3}
\def\sjnp#1#2#3 {{\em Sov. J. Nucl. Phys.} {\bf#1} (#2) #3}
\def\spj#1#2#3  {{\em Sov. Phys. JEPT} {\bf#1} (#2) #3}
\def\zp#1#2#3   {{\em Zeit. Phys.} {\bf#1} (#2) #3}
%
%
\newcommand{\clearemptydoublepage}{\newpage{\pagestyle{empty}\cleardoublepage}}
\newcommand{\scaption}[1]{\caption{\protect{\footnotesize  #1}}}
\newcommand{\proc}[2]{\mbox{$ #1 \rightarrow #2 $}}
\newcommand{\average}[1]{\mbox{$ \langle #1 \rangle $}}
\newcommand{\av}[1]{\mbox{$ \langle #1 \rangle $}}



%
%

\title{VECTOR MESON PRODUCTION AT HERA}

\author{PIERRE MARAGE 
\footnote{
On behalf of the H1 and ZEUS Collaborations.}
}

\institute{Faculty of Science, Universit\'e Libre de Bruxelles, Bd. du Triomphe\\
Brussels, B-1040,
Belgium\\
pmarage@ulb.ac.be}

\maketitle



\begin{abstract}
The rich experimental landscape of exclusive vector meson production at the high energy 
electron--proton collider HERA is reviewed, with emphasis on the transition from soft to hard 
diffraction and QCD interpretations.

\end{abstract}


\section{Introduction}	

\begin{figure}[htbp]
\begin{center}
\setlength{\unitlength}{1.0cm}
\begin{picture}(10.0,2.8)   
\put(0.0,0.0){\psfig{file=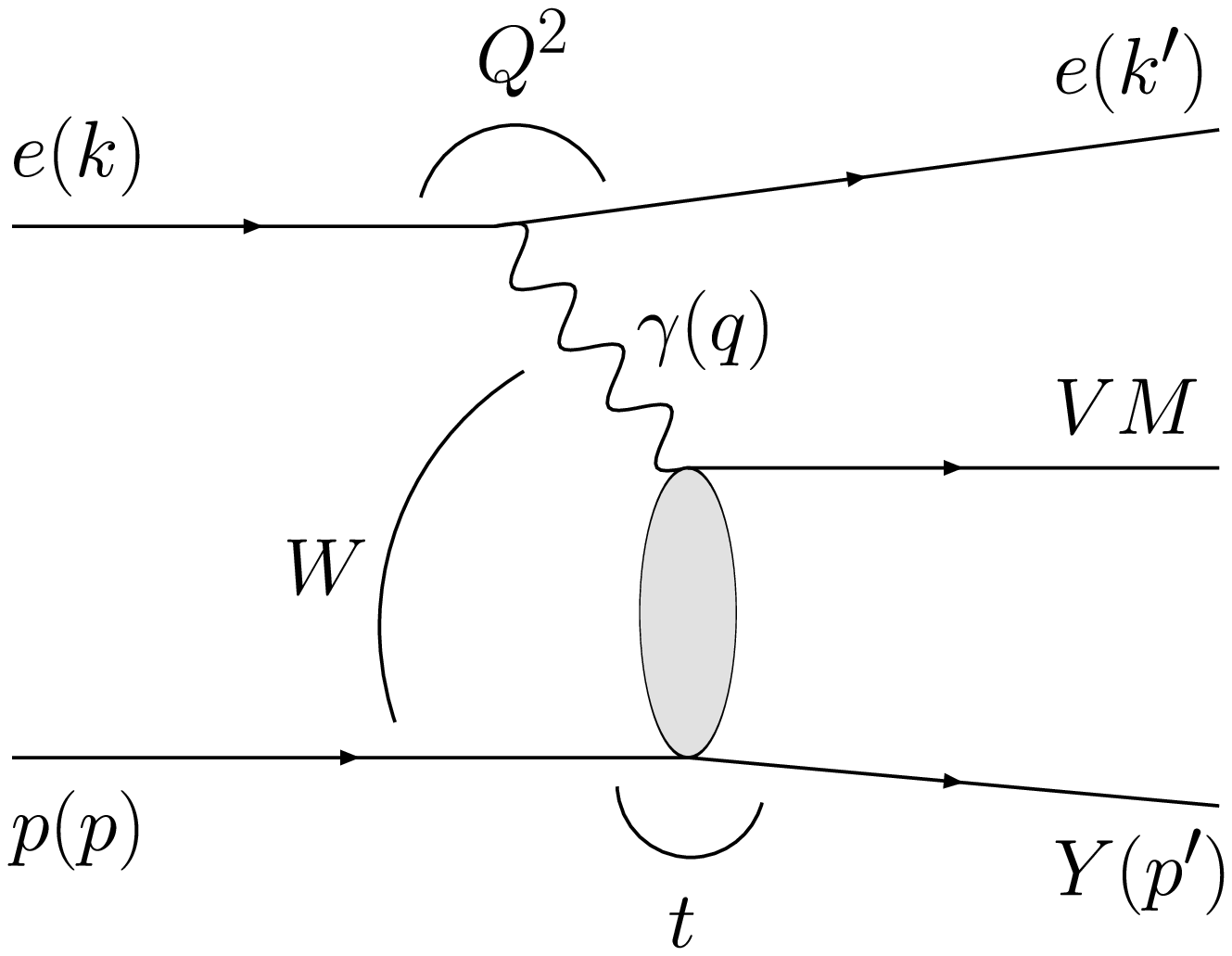,width=4.cm,height=2.8cm}}
\put(6.,0.3){\epsfig{file=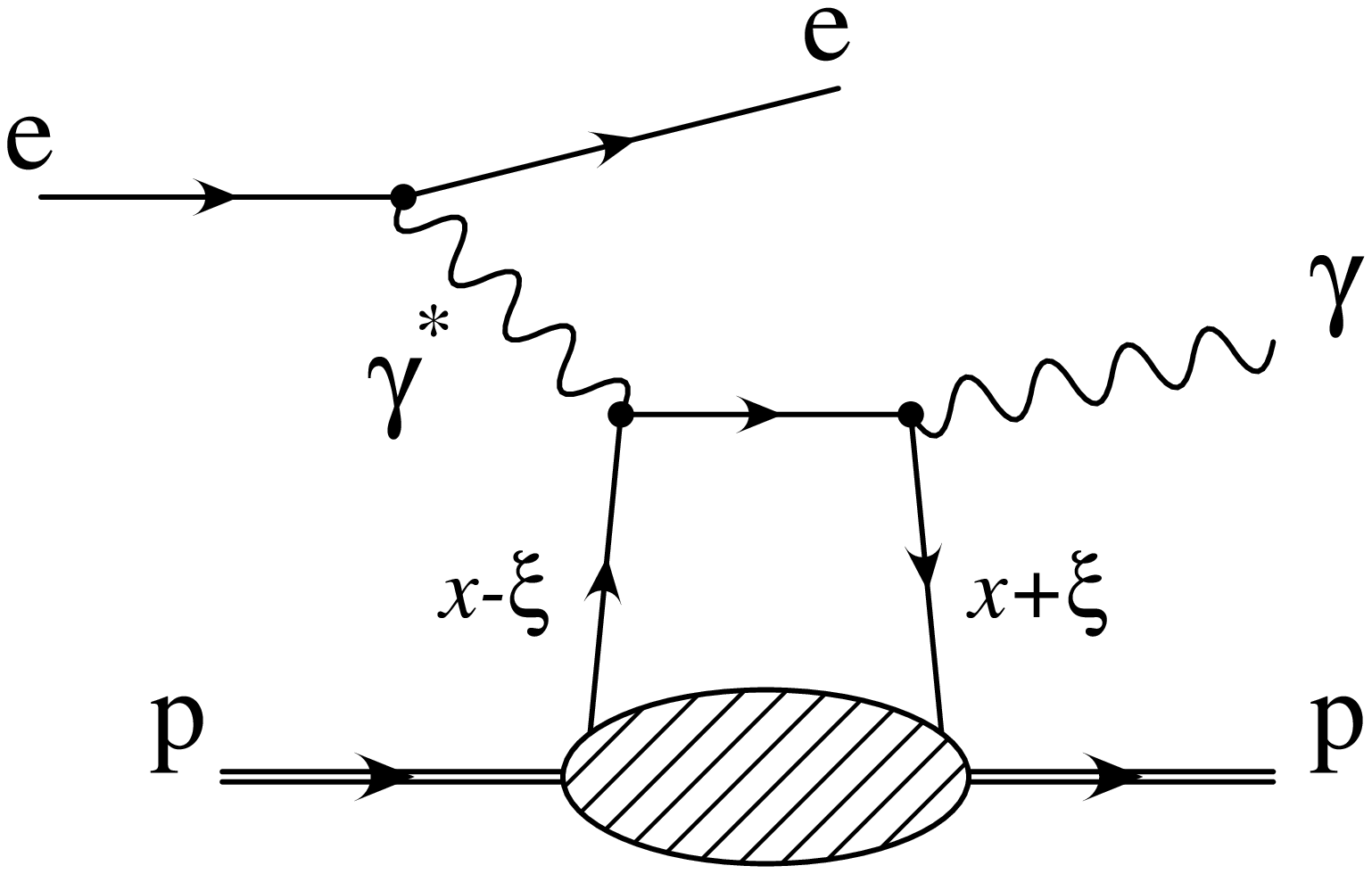,width=4.0cm,height=2.8cm}}
\end{picture}
\vspace*{-0.3cm}
\caption{(left) Diffractive vector meson electroproduction; (right) DVCS process (LO diagram).}
    \label{fig:VM}
\end{center}
\end{figure}

Vector meson (VM) production at HERA  is illustrated in Fig.~\ref{fig:VM}(left): the intermediate 
photon converts into a 
diffractively scattered $q \bar q$ pair forming a VM, while the proton remains intact 
or is diffractively excited into the system $Y$.
Here \qsq\ is the negative square of the photon four-momentum, $W$ the photon-proton center 
of mass energy ($W \simeq \qsq / x$, $x$ being the Bjorken scaling variable) and $t$ 
the square of the four-momentum transfer at the proton vertex.

The bulk of diffractive processes is governed by soft physics, which is described by Regge 
theory,
whereas the presence of a hard scale allows the use of perturbative 
calculations and the QCD understanding of diffraction.
The scale can be provided by the photon virtuality $Q$, the quark mass or the momentum
transfer $\sqrt{\modt}$. 

At large energy, VM production can be described in the proton rest frame as the 
factorisation of three processes:
the fluctuation of the virtual photon into a $q \bar q$ colour dipole~\cite{Mueller,NZ}, 
the (flavour independent) elastic (or proton dissociative) dipole-proton scattering, 
and the $q \bar q$ recombination into the final state VM.
The dipole-proton scattering is governed by the transverse size of the dipole, 
the characteristic scale of the interaction for longitudinal amplitudes of light VM and for 
heavy VM being $1/4 \qsqplmsq$.
Cross sections are thus expected to scale with this variable~\cite{FKS,ins} 
(for light VM transverse amplitudes, the scale is reduced by photon
wave function effects).

The dipole-proton scattering is modelled in perturbative QCD (pQCD) as the exchange of a 
colour singlet two-gluon system (leading order) or a BFKL ladder (LL $1/x$ approximation).
Cross sections are thus given by the square of the gluon density in the proton, which is
rapidly increasing at small $x$~\cite{ryskin,brodsky}.
The ensuing strong increase with $W$ of VM production is a characteristic feature of hard 
diffraction.
Beyond the LL $1/x$ approximation, $k_t$-unintegrated  gluon distributions obtained from 
the \qsq\ logarithmic derivative of the usual gluon distribution are used~\cite{FKS,ins,mrt,ik}.
Kinematics imply that, for large VM masses or large \qsq\ values, the longitudinal 
momenta carried by the two gluons are different~\cite{mrt2}.
At small $x$ but without requiring the presence of a hard scale, several models 
attempt at describing VM production using universal dipole-proton cross sections measured
in inclusive processes~\cite{fss,kmw,df}, often including saturation effects~\cite{GBW,munier,CGC}.

In a complementary appraoch, a collinear factorisation theorem~\cite{cfs} has been proven in 
QCD for VM longitudinal amplitudes in the deep inelastic (DIS) domain, for all values of $x$.
This approach relies on generalised parton distributions (GPD)~\cite{rev-gpd}, which take into 
account parton longitudinal correlations and intrinsic $k_t$ (see e.g. the calculations 
of~\cite{kroll}).

The studies at HERA cover the elastic and proton dissociative production of 
real photons (DVCS process) and of \rh, \om, \ph, \jpsi, $\psi(2s)$ and $\Upsilon$ mesons, 
in a large phase space domain:
$0  <  Q^2 < 90~\gevsq$, $30 < W < 300~\gev$, 
$ 0 <  |t| <  30~\gevsq$.
The photoproduction ($\qsq \simeq 0$) of light VM at small \modt\ is governed by soft 
physics, whereas heavy VM production is hard and pQCD calculable.
The transition from soft to hard diffraction is studied with light VM electroproduction 
($Q^2 >$ a few~\gevsq).
Large \modt\ studies give specific access to hard, BFKL-type processes.

Cross sections, expressed in terms of \gstarp\ scattering,
are measured as a function of \qsq, $W$ and $t$, and angular distributions 
give access to spin density matrix elements and helicity amplitudes.

\section{From soft to hard diffraction: the mass scale}	

\begin{figure}[htbp]
\begin{center}
\setlength{\unitlength}{1.0cm}
\begin{picture}(7.,7.0)   
\put(0.0,0.0){\epsfig{file=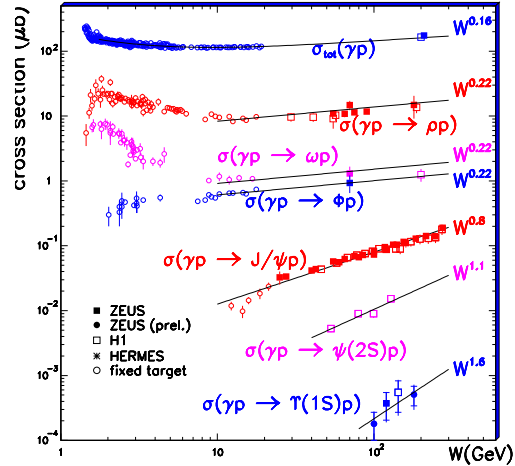,width=7.0cm,height=7.0cm}}
\end{picture}
\vspace*{-0.3cm}
\caption{VM photoproduction cross sections (from\protect\cite{z-upsilon}).}
    \label{fig:xsect}
\end{center}
\end{figure}

Total cross sections for VM photoproduction are presented as a function of the proton-photon 
center of mass energy $W$ in Fig.~\ref{fig:xsect}.
They are parametrized, in a Regge inspired form, as $\sigma (\gstarp)  \propto W^{\delta}$
with $\delta =  4 \ ( \alpom ( \langle t \rangle )  - 1)$ and 
$\alpom(t) = \alpom(0) + \alp \cdot t$.

The soft pomeron trajectory measured in hadron-hadron interactions with intercept 
$\alpom(0) \simeq 1.1$ and slope $\alp \simeq 0.25~\gevsqm$ describes well the
weak energy dependence of the total photoproduction cross section and of light 
vector meson production
(\rh~\cite{z-rho-photoprod,z-VM-photoprod,olsson},
\om~\cite{z-omega-photoprod}, 
\ph~\cite{z-VM-photoprod,z-phi-photoprod}).

The energy dependence of heavy vector mesons 
(\jpsi~\cite{z-jpsi-photoprod,h1-jpsi-hera1}, 
$\psi (2s)$~\cite{h1-psi2s}, 
$\Upsilon$~\cite{z-upsilon,h1-upsilon})
is strikingly different, with much larger values of $\delta$ and $\alpom(0)$.
This hard behaviour confirms QCD expectations, the hard scale being provided by the 
heavy quark mass.
For $\Upsilon$ production, the strong rise of the cross section is described by 
calculations using GPD's, which take into account the contributions of the real part of the amplitudes 
and the strong kinematical constraints due to the large VM mass.

The \modt\ distributions, which are exponentially falling for $\modt\ \lapprox\ 1~\gevsq$,
are also strikingly different for light and heavy VM: 
the exponential $b$ slopes values are around $8-10~\gevsqm$ for \rh\ 
photoproduction and around $4-5~\gevsqm$ for \jpsi.
Since the slope results from the convolution of the transverse sizes of the proton, 
of the $q \bar q$ dipole and of the exchanged system, the shallower 
\modt\ distributions for heavy VM supports the dominance of small dipoles, related to the
large quark mass. 

The slope \alp\ of the effective Regge trajectory, which measures the correlation between 
the $W$ and $t$ dependences, is smaller for \jpsi\ photoproduction 
($\alp \simeq 0.12-0.16~\gevsqm$) than for soft hadronic diffraction, which 
may be another signature of hard diffraction.

\section{From soft to hard diffraction: the {\boldmath \qsq} scale}	

\begin{figure}[htbp]
\begin{center}
\setlength{\unitlength}{1.0cm}
\begin{picture}(12.0,6.0)   
\put(0.0,0.0){\epsfig{file=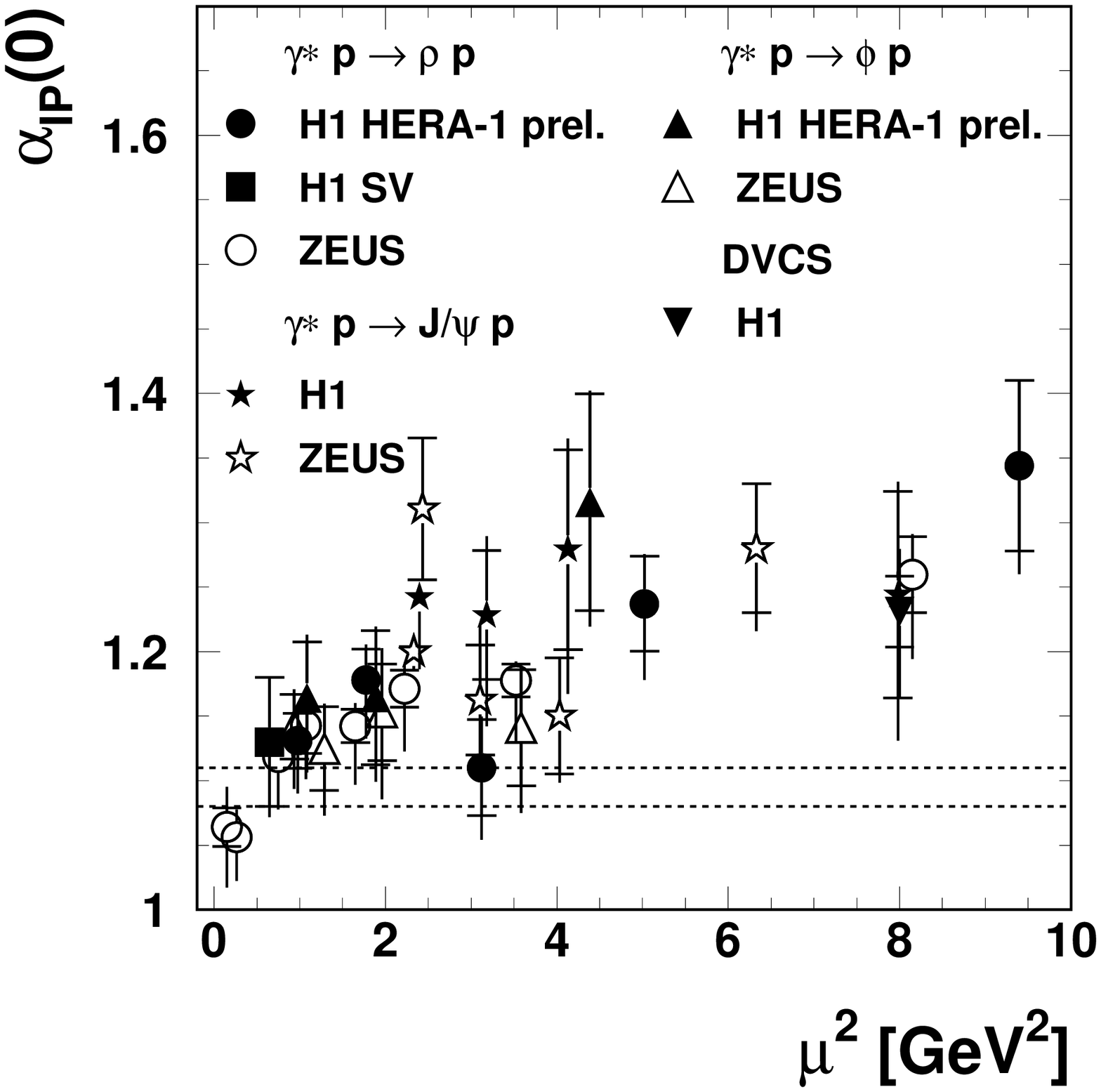,height=6.0cm,width=6.0cm}}
\put(6.0,0.0){\epsfig{file=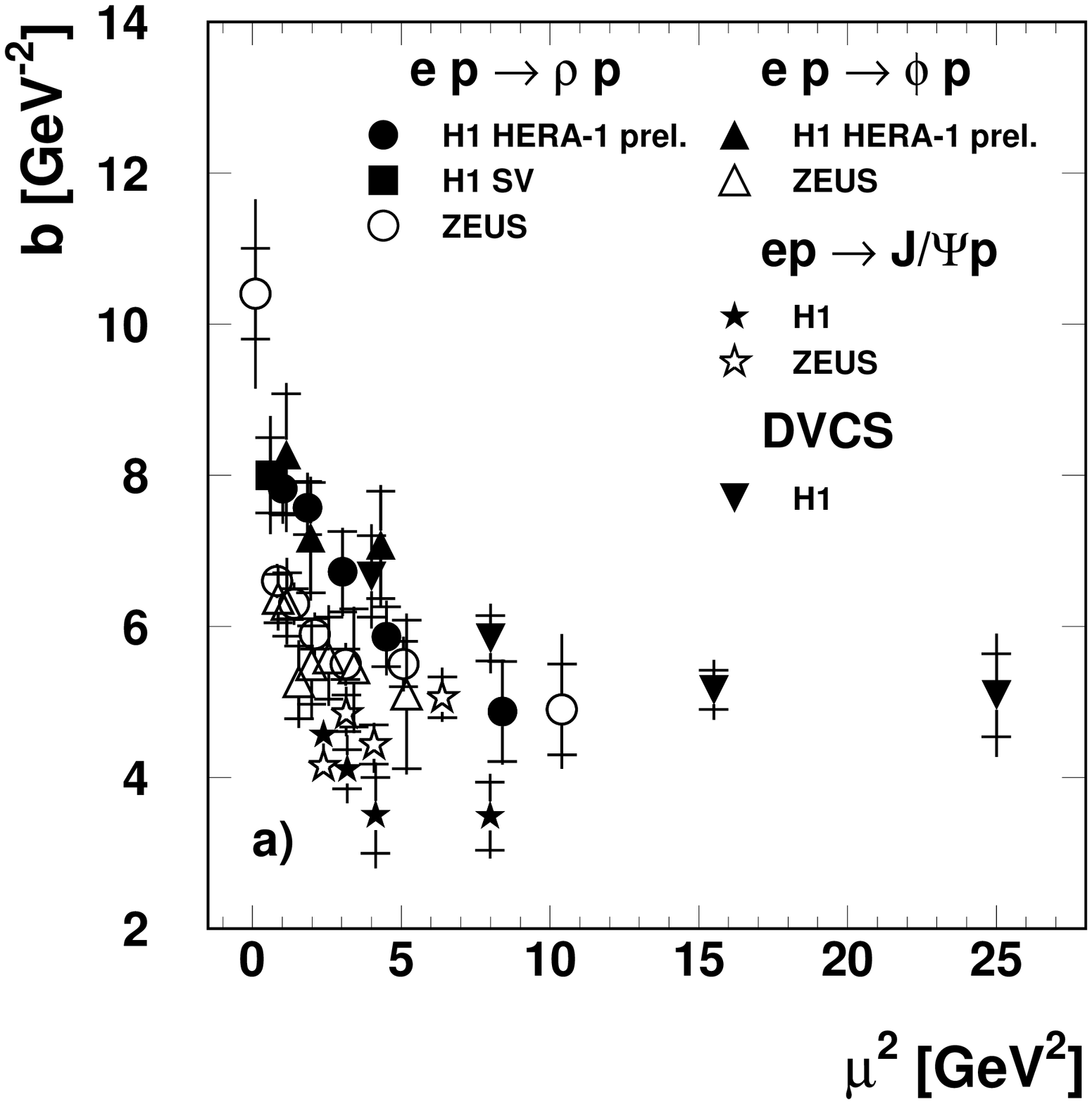,height=6.0cm,width=6.0cm}}
\put(7.3,1.3){\epsfig{file=figures/whitebox.eps,width=0.4cm,height=0.4cm}}
\end{picture}
\vspace*{-0.3cm}
\caption{
(left) Evolution with the scale $\mu^2$ of the parameter \alpom(0), describing the $W$
dependence of DVCS and of VM production 
(the dashed lines show typical values for soft diffraction);
(right) $b$ slope parameters of the \modt\ distributions.
}
\label{fig:alphapom0-b_f_qsq}
\end{center}
\end{figure}

\paragraph {Deeply virtual Compton scattering} 
The DVCS process in electroproduction, $e + p \to e + \gamma + p$~\cite{z-dvcs,h1-dvcs},
is described at leading order in Fig.~\ref{fig:VM}(right), with the use of GPD's taken at the 
scale $\mu^2= Q^2$.
The $W$ and $t$ dependences are significantly harder than for photoproduction of  light VM, 
and similar to those of heavy VM at the scale $\mu^2 = 1/4 \qsqplmsq$
(see Fig.~\ref{fig:alphapom0-b_f_qsq}).

\paragraph {Electroproduction of \rh\ and \ph\ mesons}
The \qsq\ dependences of light (\rh~\cite{z-rho,h1-hera1} and \ph~\cite{h1-hera1,z-phi}) VM 
electroproduction are roughly described by power laws
$1 / Q^{2n}$, with $n \simeq 2.-2.5$.
Several effects modify the simple $n = 3$ dependence expected in a naive two-gluon approach for
the longitudinal cross sections, in particular the \qsq\ dependence of the gluon 
distribution at small $x$, the \qsq\ dependence of the ratio 
$R = \sigma_L / \sigma_T$ of the longitudinal to transverse cross sections, and the delayed
approach to hard diffraction of the latter.
VM WF effects may also play a role.

\begin{figure}[htbp]
\begin{center}
\setlength{\unitlength}{1.0cm}
\begin{picture}(12.0,6.0)   
\put(0.0,0.0){\epsfig{file= 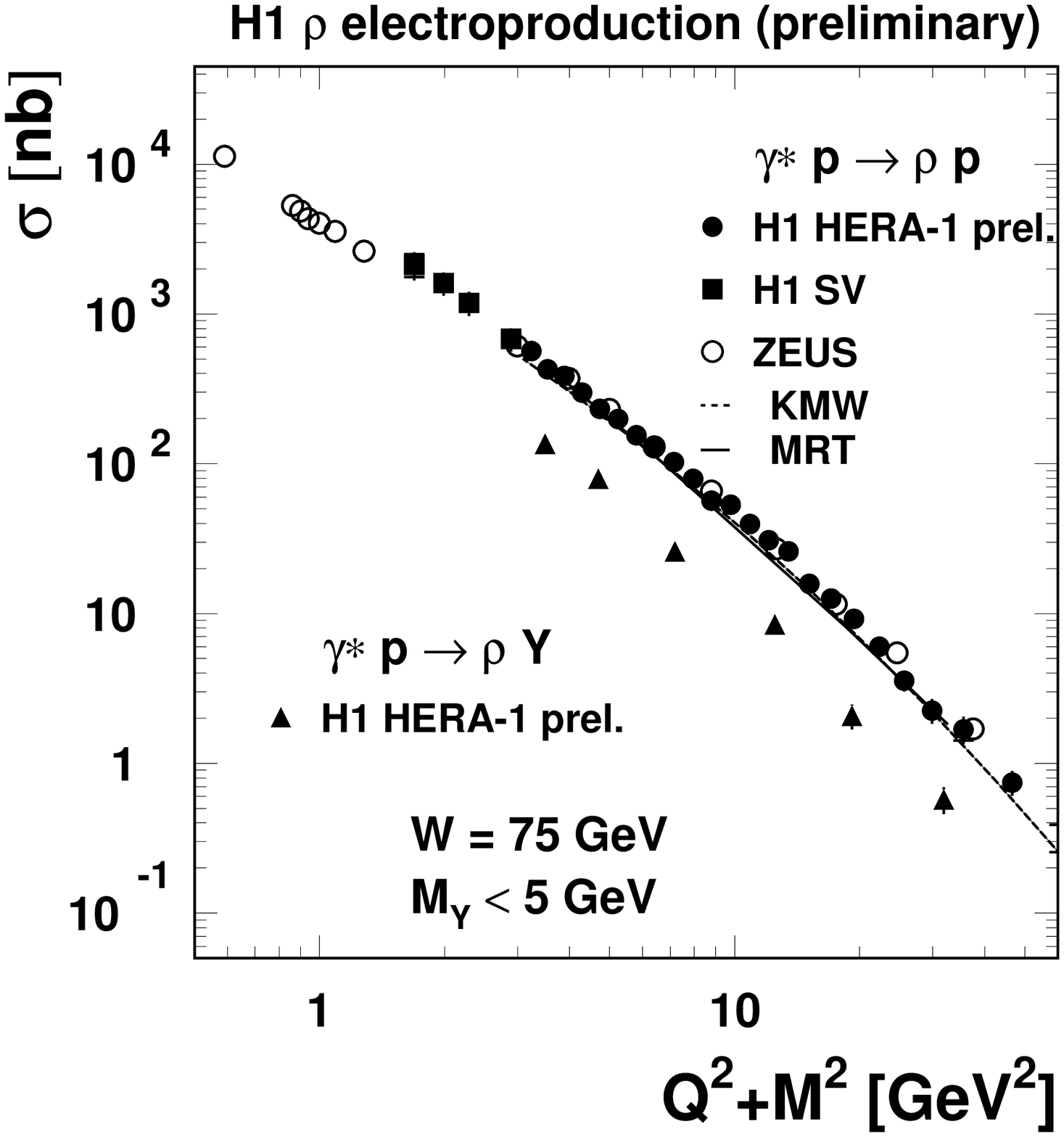,height=6.0cm,width=6.0cm}}
\put(1.3,1.3){\epsfig{file=figures/whitebox.eps,width=0.4cm,height=0.4cm}}
\put(1.0,5.4){\epsfig{file=figures/whitebox.eps,width=5.0cm,height=0.6cm}}
\put(6.0,0.5){\epsfig{file=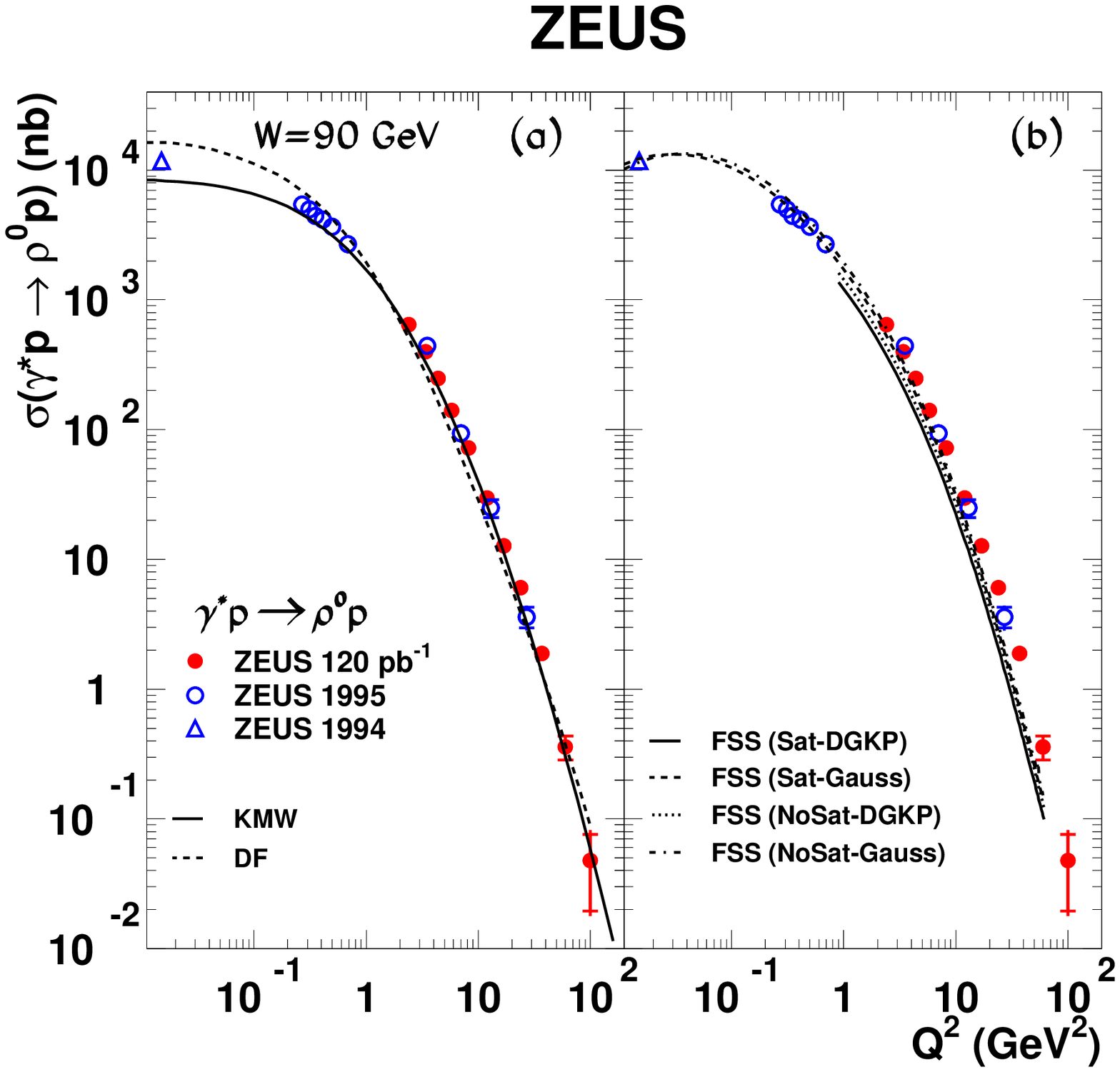,width=6.0cm,height=5.5cm}}
\put(6.0,5.5){\epsfig{file=figures/whitebox.eps,width=6.0cm,height=0.5cm}}
\end{picture}
\vspace*{-0.3cm}
\caption{
\qsq\ dependence of \rh\ production cross section 
(from\protect\cite{z-rho,h1-hera1}).}
\label{fig:light-qsq}
\end{center}
\end{figure}

A good description of the \qsq\ dependence is given by pQCD approaches using 
$k_t$-unintegrated gluon distributions (MRT~\cite{mrt}) or GPD~\cite{kroll} (not shown).
Dipole models using different saturation and wave function parametrizations 
(FSS~\cite{fss}, KMW~\cite{kmw}, DF~\cite{df}) attempt with reasonable success at 
describing VM production over the full \qsq\ range, including photoproduction.

The $W$ dependences for \rh\ and \ph\ mesons
become harder as \qsq\ increases, with \alpom(0) reaching values similar to those of
\jpsi\ photoproduction, $\alpom(0) \simeq 1.3$, for $1/4(\qsq + M^2) \gapprox\ 3-5~\gevsq$ 
(Fig.~\ref{fig:alphapom0-b_f_qsq} left).
Similarly, the $t$ dependence of light VM production becomes shallower has \qsq\ increases,
which provides another indication of the role of small dipoles at high \qsq.
The slope of the effective Regge trajectory has a value smaller than for the soft pomeron,
 $\alp \simeq 0.10 - 0.15~\gevsqm$.

\paragraph {\jpsi\ electroproduction}
For heavy VM electroproduction~\cite{h1-jpsi-hera1,z-jpsi-elprod}, 
\qsq\ provides a second hard scale, in addition to the quark mass.
The energy dependence, parametrized with $\alpom(0) \simeq 1.2-1.3$, 
and the $t$ dependence, with $b \simeq 4-5~\gevsqm$, do not depend significantly on \qsq\ 
(Fig.~\ref{fig:alphapom0-b_f_qsq}).
Indications are found that the effective trajectory slope is reduced, with 
$\alp = 0.07 \pm 0.05~\gevsqm$.


\section {Universality; factorisation}

\begin{figure}[htbp]
\begin{center}
\setlength{\unitlength}{1.0cm}
\begin{picture}(12.0,6.0)   
\put(0.0,0.0){\epsfig{file= 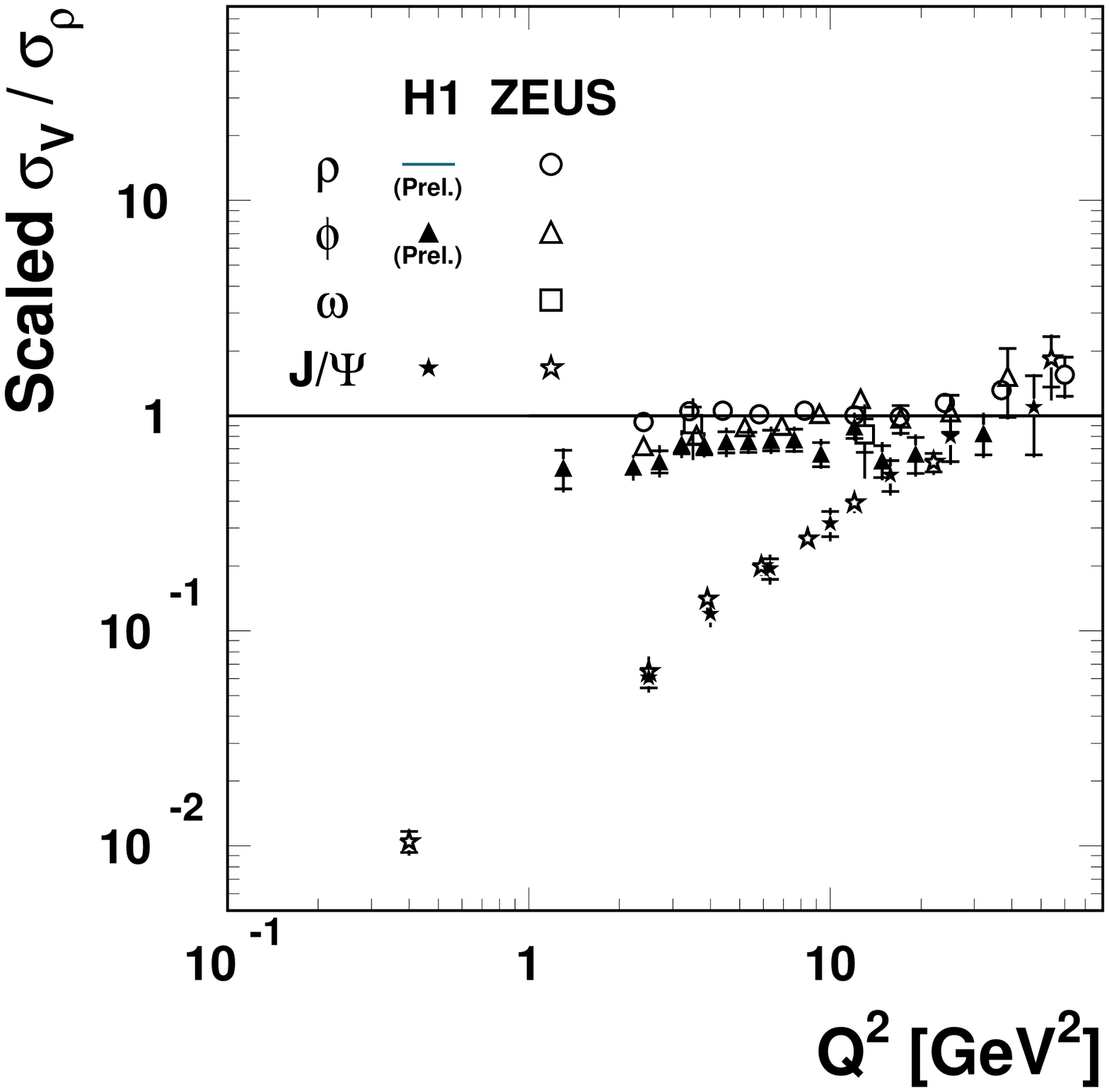,height=6.0cm,width=6.0cm}}
\put(6.0,0.0){\epsfig{file= 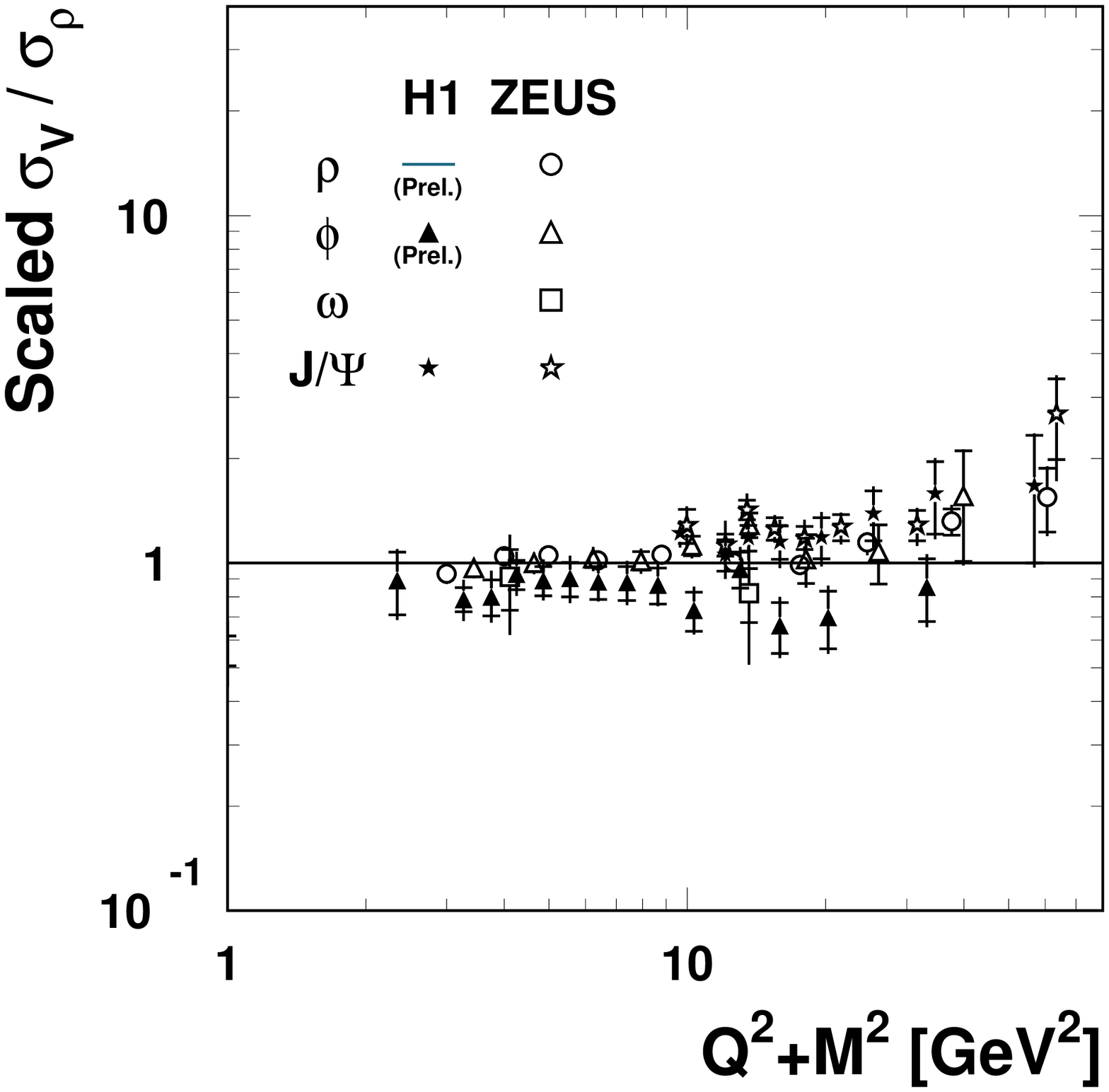,height=6.0cm,width=6.0cm}}
\end{picture}
\vspace*{-0.3cm}
\caption{
Ratios of VM elastic production cross sections, divided
by the (preliminary)  \rh\ measurements of H1 (shown as the line at 1) 
and scaled according to the quark charge content (\rh\ : \ph\ : \jpsi = 8 : 2 : 9),
plotted as a function of \qsq\ (left)  and \qsqplmsq\ (right).
}
\label{fig:SU5_sigma_f_qsqplmsq}
\end{center}
\end{figure}

Figure~\ref{fig:SU5_sigma_f_qsqplmsq} presents elastic production cross sections
for \rh, \om, \ph\ and \jpsi, divided by the preliminary \rh\ measurement of H1 and scaled by 
the VM charge content, as a function of \qsq\ (left) and as a function of \qsqplmsq\ 
(right).
Whereas they differ by orders of magnitude at small \qsq,
the ratios are close to 1 when plotted as a function of the scaling variable \qsqplmsq.
This supports the idea that the cross sections are essentially determined by the dipole size,
with little flavour dependence (WF effects should however be taken into account for detailed 
comparisons).

Universality is also supported by the common $W$ and $t$ 
dependences of the different VM, for the scale $\mu^2 \ \gapprox\ 3-5~\gevsq$
(Fig.~\ref{fig:alphapom0-b_f_qsq}, \alpom(0) and $b$ parameters). 

The ratio of the elastic to proton dissociation cross sections for $t = 0$ are independent 
of \qsq~\cite{h1-hera1},
which supports the factorisation of the processes at the photon and proton vertices.

Factorisation is confirmed~\cite{h1-hera1} by the \qsq\ independence of the difference between the longitudinal and
transverse slopes, $b_L - b_T$, a quantity which is also found to be consistent for different VM
and expected to be related to the proton size.

\section{Large $t$ scattering}	

A hard scale for diffractive VM production is also provided by $|t|$.
For large \modt\ \jpsi\ production, a hard scale is present at both ends of the exchange, and no strong $k_t$ 
ordering along the ladder is expected, which is typical for BFKL calculations.

\begin{figure}[htbp]
\begin{center}
\setlength{\unitlength}{1.0cm}
\begin{picture}(10.0,6.0)   
\put(0.5,0.0){\epsfig{file= 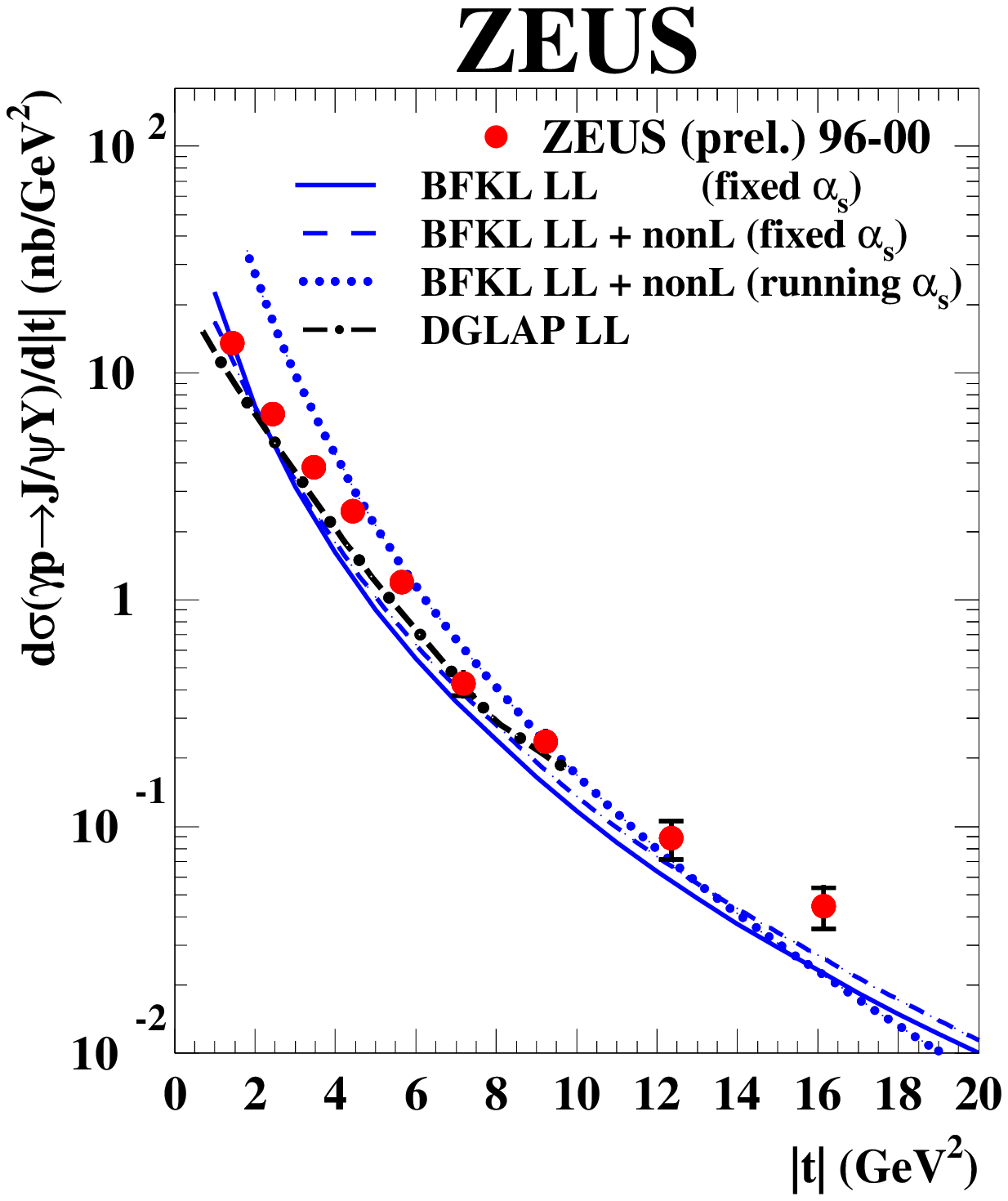,height=5.8cm,width=5.8cm}}
\put(0.0,4.65){\epsfig{file=figures/whitebox.eps,width=7.0cm,height=0.3cm}}
\put(5.5,0.6){\epsfig{file= 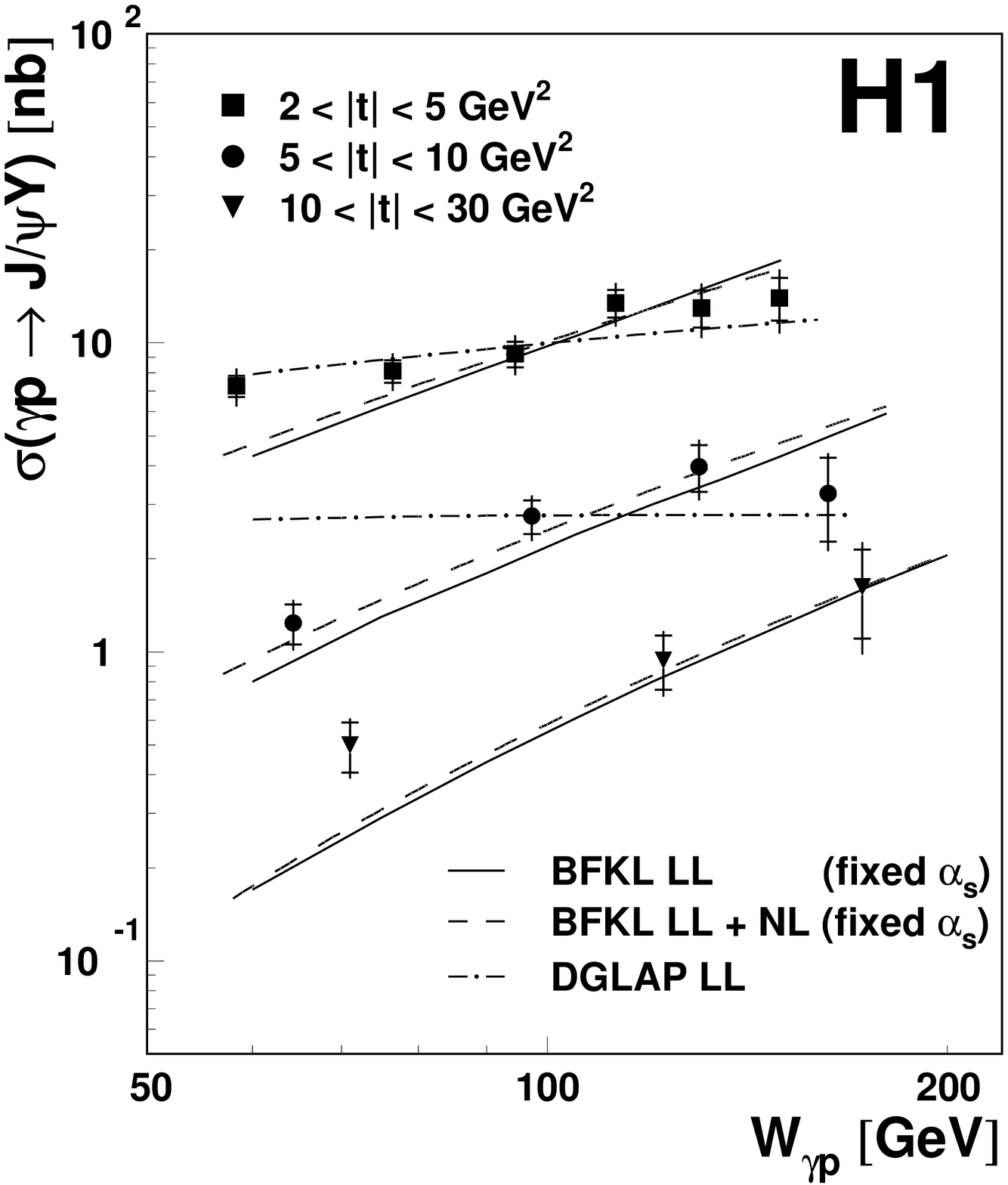,height=4.5cm,width=4.5cm}}
\end{picture}
\vspace*{-0.5cm}
\caption{
$t$ (left) and $W$ (right) dependences of \jpsi\ production with $\modt > 2~\gevsq$ 
(from\protect\cite{z-high-t-prel,h1-jpsi-photoprod-large-t}).  
}
\label{fig:large-t}
\end{center}
\end{figure}

In these processes, dominated by proton dissociation, the $t$ dependence of the cross 
sections obeys power laws, both for \rh~\cite{z-high-t,h1-rho-photoprod-large-t} and 
\jpsi~\cite{z-high-t-prel,h1-jpsi-photoprod-large-t} photoproduction.
They are well described by BFKL-based pQCD calculations with fixed $\alpha_s$ and also, for 
$\modt\ \lapprox\ m_\psi^2$, using the DGLAP evolution (Fig.~\ref{fig:large-t} left).
BFKL calculations also describe the $W$ evolution (Fig.~\ref{fig:large-t} right),
at variance with DGLAP, but does not describe well the spin density matrix elements.

For \jpsi\ photoproduction with $\modt\ \gapprox\ 2~\gevsq$, the slope of the effective Regge 
trajectory tends to be negative.

\section{Helicity amplitudes}	

\begin{figure}[htbp]
\begin{center}
\setlength{\unitlength}{1.0cm}
\begin{picture}(13.,9.0)   
\put(-0.2,0.0){\epsfig{file=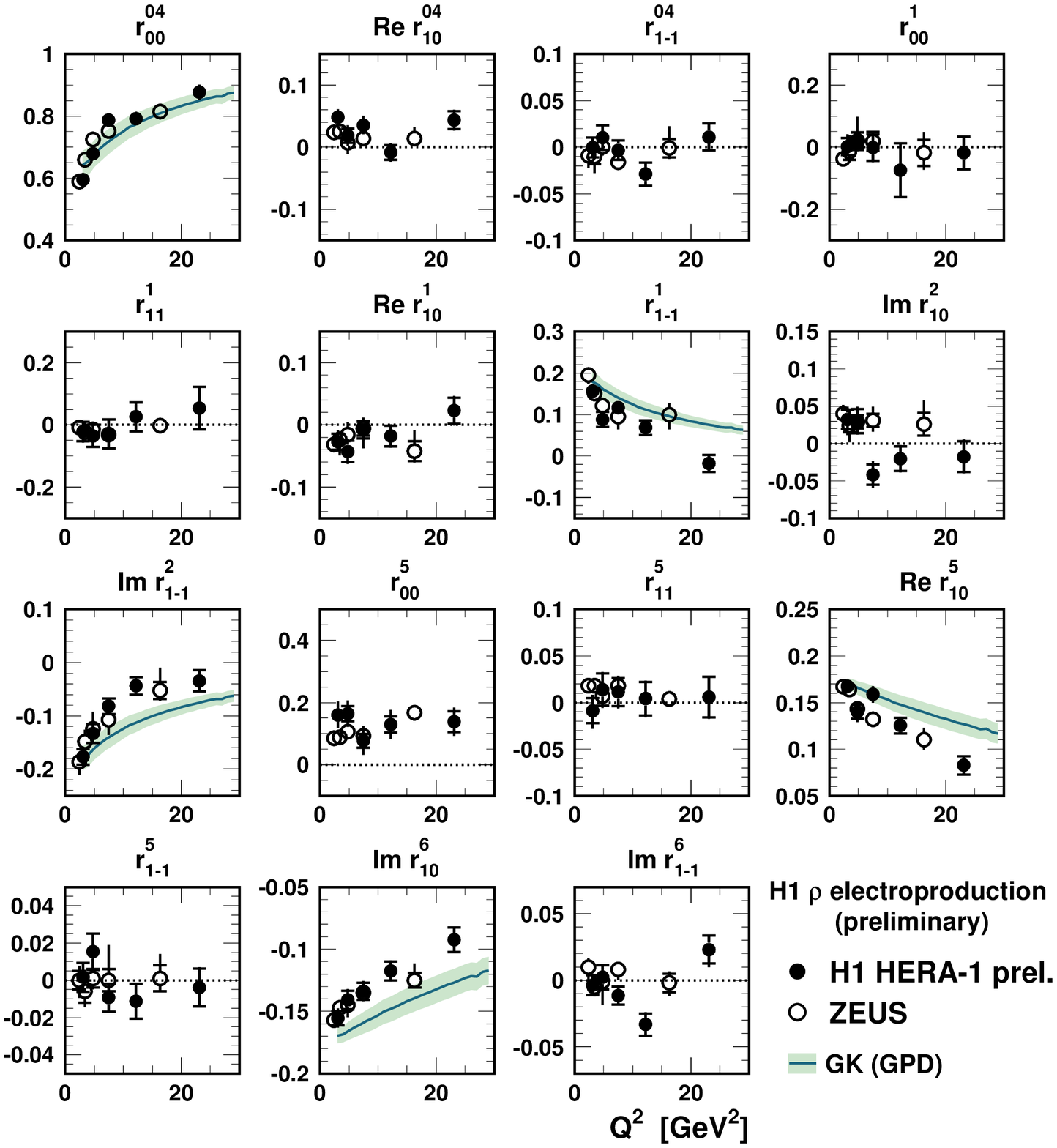,width=9.0cm,height=9.0cm}}
\put(9.2,5.8){\epsfig{file= 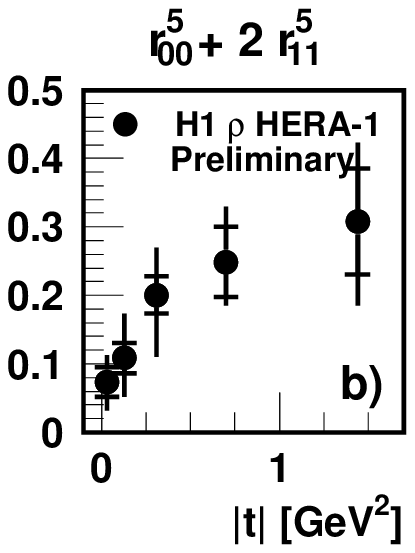,height=3.cm,width=3.cm}}
\put(8.7,0.2){\epsfig{file= 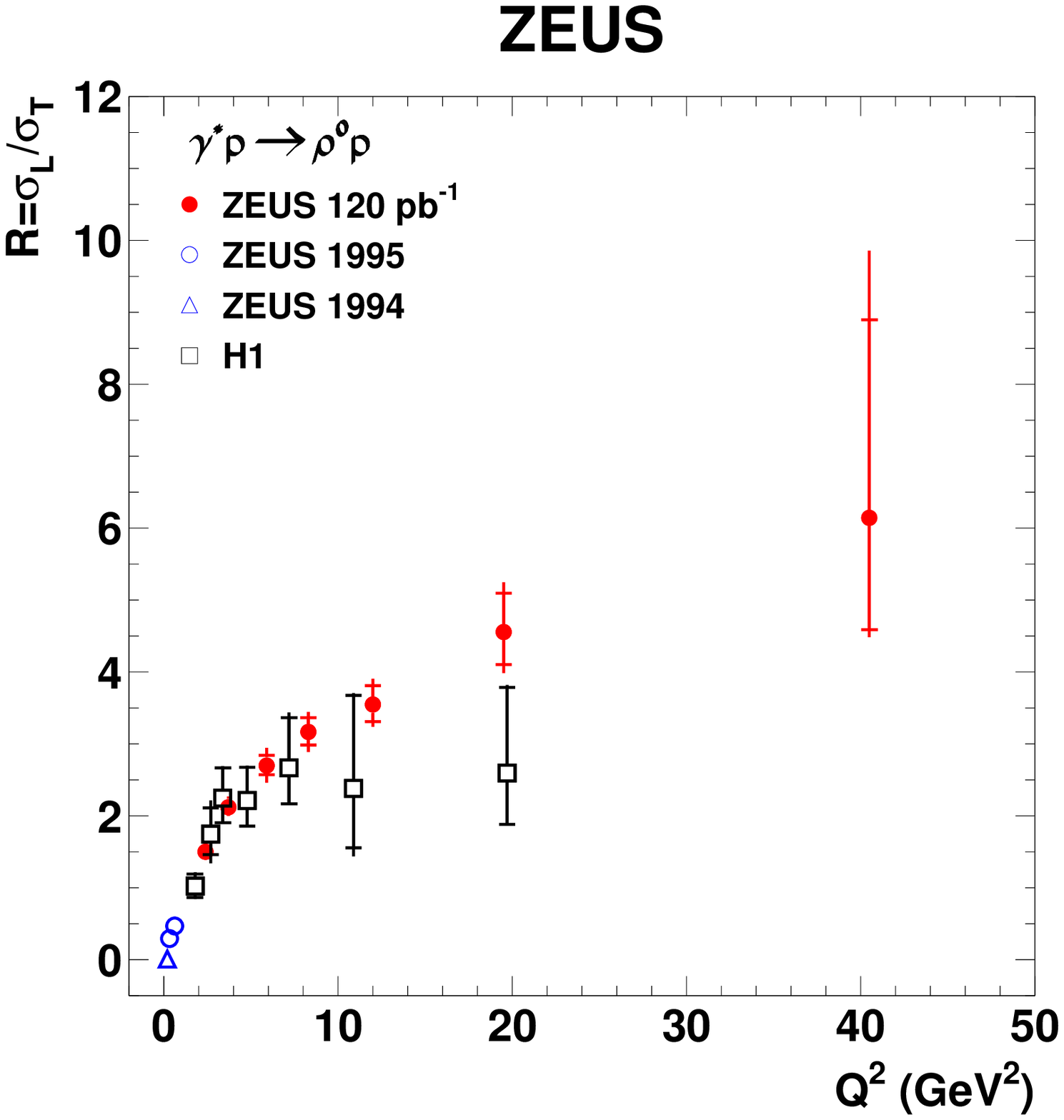,height=4.5cm,width=4.5cm}}
\put(8.7,4.32){\epsfig{file=figures/whitebox.eps,width=4.5cm,height=0.5cm}}
\end{picture}
\vspace*{-0.3cm}
\caption{(left) Spin density matrix elements for \rh\ meson production, as a function of \qsq;
(right up) $t$ dependence of the matrix element combination \rfivecomb;
(right bottom) \qsq\ dependence of the ratio $R = \sigma_L / \sigma_T$ of the 
longitudinal to transverse cross sections
(from\protect\cite{z-rho,h1-hera1}).  
}
\label{fig:sdme}
\end{center}
\end{figure}

The measurement of angular distributions gives access to spin density matrix elements, which
are related to the helicity amplitudes.
Figure~\ref{fig:sdme} (left) presents the 15 \rh\ matrix elements as a function of \qsq, 
compared to predictions of a GPD model (GK~\cite{kroll}).
Other models also describe qualitatively the data, but not quantitatively, e.g. the $k_t$-unintegrated 
gluon model (INS~\cite{ins}, with sensitivity to parametrizations of the \rh\ WF).

The $s$-channel helicity conserving (SCHC) amplitudes are 
dominant, with the longitudinal amplitude larger than the transverse one
(Fig.~\ref{fig:sdme} right, bottom).
A significant contribution of the transverse to longitudinal helicity flip amplitude, 
increasing with \modt, is observed 
(Fig.~\ref{fig:sdme} right, up).
These features are qualitatively understood in pQCD~\cite{ik}.

\section{Conclusions}	

Studies of DVCS and VM production at HERA provide a rich and varied
field for the QCD understanding of diffraction and the testing of calculations, over a 
large kinematic domain.
Whereas soft diffraction, similar to hadronic interactions, dominates light VM photoproduction,
typical features of hard diffraction arise with the developments of hard scales provided
by \qsq, the quark mass or \modt, 
in particular hard $W$ and shallow $t$ dependences. 
Calculations based on pQCD, notably using GPD approaches, and models using 
universal dipole cross sections describe the data relatively well.
Angular distributions and the measurement of spin density matrix elements give a detailed
access to the polarisation amplitudes, which is also understood in QCD.

\section*{Acknowledgments}

It is a pleasure to thank the numerous colleagues, theorists as well as experimentalists from
H1 and ZEUS, with whom interesting and vivid discussions over the years have made 
the study of diffraction at HERA so enjoyable.


\end{document}